\documentclass[twocolumn]{article}

\usepackage{arxiv}

\usepackage{cite}
\usepackage{amsmath,amssymb,amsfonts}
\usepackage{algorithmic}
\usepackage{graphicx}
\usepackage{textcomp}
\usepackage{textgreek}
\usepackage{soul,color}
\usepackage{booktabs}
\usepackage{multirow}
\usepackage{caption}
\usepackage{subcaption}
\usepackage{hyperref}

\title{Fighting the scanner effect in brain MRI segmentation with a progressive level-of-detail network trained on multi-site data}

\author{
Michele Svanera\thanks{Corresponding author: Michele.Svanera at glasgow.ac.uk - $^\diamond$ Shared authorship.}\\
Center for Cognitive Neuroimaging\\
School of Psychology \& Neuroscience\\
University of Glasgow (UK)
\And
Mattia Savardi\\
Department of Medical and Surgical Specialties, \\
Radiological Sciences, and Public Health\\
University of Brescia (Italy)
\And
Alberto Signoroni\\
Department of Medical and Surgical Specialties, \\
Radiological Sciences, and Public Health\\
University of Brescia (Italy)
\And
Sergio Benini $^\diamond$\\
Department of Information Engineering\\
University of Brescia (Italy)
\And
Lars Muckli $^\diamond$\\
Center for Cognitive Neuroimaging\\ 
School of Psychology \& Neuroscience\\
University of Glasgow (UK)
}

\begin{document}
\maketitle
\vspace{-10cm}

\twocolumn[
\begin{abstract}

Many clinical and research studies of the human brain require an accurate structural MRI segmentation.
While traditional atlas-based methods can be applied to volumes from any acquisition site, recent deep learning algorithms ensure very high accuracy only when tested on data from the same sites exploited in training (\textit{i.e.}, internal data).
The performance degradation experienced on external data (\textit{i.e.}, unseen volumes from unseen sites) is due to the inter-site variabilities in intensity distributions induced by different MR scanner models, acquisition parameters, and unique artefacts.
To mitigate this site-dependency, often referred to as the \textit{scanner effect}, we propose \texttt{LOD-Brain}, a 3D convolutional neural network with progressive levels-of-detail (LOD) able to segment brain data from any site.
Coarser network levels are responsible to learn a robust anatomical prior useful for identifying brain structures and their locations, while finer levels refine the model to handle site-specific intensity distributions and anatomical variations.
We ensure robustness across sites by training the model on an unprecedented rich dataset aggregating data from open repositories: almost 27,000 T1w volumes from around 160 acquisition sites, at 1.5 - 3T, from a population spanning from 8 to 90 years old.  
Extensive tests demonstrate that \texttt{LOD-Brain} produces state-of-the-art results, with no significant difference in performance between internal and external sites, and robust to challenging anatomical variations.
Its portability opens the way for large scale application across different healthcare institutions, patient populations, and imaging technology manufacturers. Code, model, and demo are available at the \href{https://rocknroll87q.github.io/LOD-Brain/}{project website}.

\end{abstract}
\bigskip
]

\section{Introduction}
\label{sec:introduction}

Brain structure segmentation in magnetic resonance imaging (MRI) plays a pivotal role in both research and clinical routines for assessing and monitoring brain morphology, volumetry, and connectivity, in both normal and pathophysiological conditions. 
As more and more studies analyse data derived from thousands of MRI brain scans \cite{bethlehem2022brain}, there is a growing need for tools able to perform automatic, fast, and reliable segmentation of brain structures, with benefits on downstream research and clinical studies in terms of accuracy, statistical power, and reproducibility of findings.

Well-established segmentation methods in neuroimaging, such as FreeSurfer \cite{fischlFreeSurfer2012a} and FSL~\cite{jenkinson2012fsl}, exploit one or more atlases \textit{i.e.}, reference volumes and their manual trusted segmentation: first the target is registered with the reference volume, then the anatomical prior knowledge from the manual segmentation is transferred to the target volume \cite{yaakub2020brain}.
Although computationally expensive and slow, these methods easily adapt to images from different scanners or acquired by means of different sequences.

Recently, achievements in deep learning (DL) methods applied to automatic brain MRI segmentation \cite{akkusDeepLearningBrain2017a} such as DeepNat \cite{wachingerDeepNATDeepConvolutional2018}, QuickNat \cite{royQuickNATFullyConvolutional2019}, and CEREBRUM~\cite{bontempi2020cerebrum,svanera2021cerebrum} have made remarkable progress in competing  with the reliability offered by atlas-based segmentation methods.
However, most DL methods usually include, for both training and testing, only MRI volumes collected from a single or few centres with almost homogeneous characteristics in terms of image statistics, acquisition parameters, and artefacts. 
Consequently, when challenged on external data \textit{i.e.}, unseen volumes from unseens sites, DL methods face the so-called \textit{scanner effect}, a drop in performance on handling the data variability originated by different MRI site acquisitions.
This mismatch between the distributions of internal and external data, which is common in MRI (see \textit{e.g.,} the competitions in ~\cite{sun2021multi-infant, campello2021multi}) is a problem more broadly known as \textit{distribution shift} \cite{wiles2021fine}.
Some researchers in brain segmentation propose to tackle it by applying aggressive data augmentation \cite{zhao2019data} or harmonisation \cite{beer2020longitudinal}, by using domain adaptation or randomisation \cite{billot2021synthseg}, or by generating synthetic data with the needed variations     \cite{shin2018medical}.
Despite achieving good robustness on a wide range of MRI contrasts and resolutions, these approaches keep showing limitations in matching statistics of real data distributions, struggling with morphological variabilities and atypical scanner artefacts.

In order to handle inter-site diversity, none of the existing DL solutions builds on the idea of generating the equivalent of an anatomical \textit{brain prior}, for example exploiting volumes from multiple sites.
Given the current availability of open datasets, a concrete opportunity for improving the model portability is in fact training a model directly on out-of-the-scanner data coming from multiple sites, to cover different vendors, resolutions, slice thickness, participant demographics, and pathological conditions.
Previous approaches to multi-site learning for segmentation in different medical imaging domains show, on the one hand, that these methods help generalising on external data.
On the other hand, they often perform worse on internal ones (\textit{i.e.}, unseen volumes from sites included in the training set) \cite{styner2002multisite}.
This apparently contradictory situation has been observed also in other medical image analysis tasks \cite{zech2018variable}, reinforcing the concept that effective learning from multiple sources is highly challenging and can introduce unexpected performance drops.  

To exploit the informative innovation carried by such multi-source data, dedicated architectural solutions should be designed.
An effective method should be able to integrate anatomical knowledge acquired by a large number of volumes into a robust anatomical brain prior. 
Additionally, this solution should handle the high degree of variability that characterises data from different sites and scanner vendors.

\subsection{Main contributions}

\begin{figure}[h]
     \centering
         \centering
         \includegraphics[width=0.95\columnwidth]{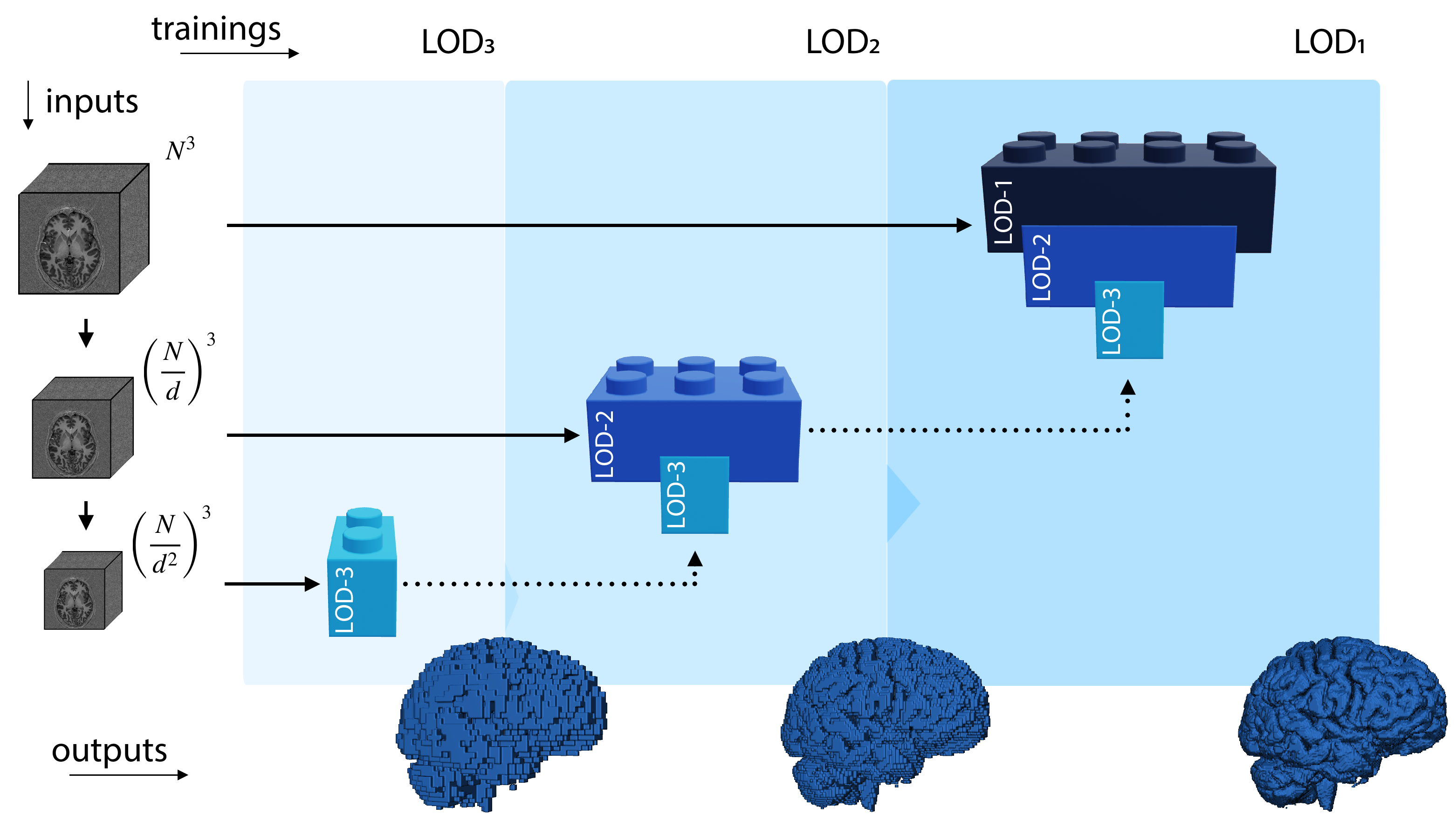}
        \caption{LOD-Brain is a level-of-detail (LOD) network, where each LOD is a U-net which processes 3D brain multi-data at a different scale. Lower levels learn a coarse and site-independent brain representation, while superior ones incorporate the learnt spatial context, and refine segmentation masks at finer scales. Examples of outputs (grey matter renderings) at different LODs are shown in blue at the bottom.}
        \label{fig:framework}
\end{figure}

We here present \texttt{LOD-Brain}, a progressive level-of-detail network for training a robust brain MRI segmentation model from a huge variety of multi-site and multi-vendor data.
\texttt{LOD-Brain} architecture is organised on multiple levels of detail (LOD), as shown in Fig.~\ref{fig:framework}.
Each level is a convolutional neural network (CNN) which processes 3D brain data at a different scale obtained via progressively down-sampling the input volume.
Thanks to the rich variability of brain samples coming from 70 datasets from different MRI acquisition sites, the proposed architecture learns, at lower levels, a robust brain anatomical prior.
Concurrently, higher levels handle site-specific intensity distributions and scanner artefacts.
Through inter-level connections between networks and a bottom-up training procedure, such architecture integrates contributions from all levels to produce an accurate and fast segmentation.

\texttt{LOD-Brain} shows outstanding generalisation capabilities, as it performs better than other state-of-the-art solutions on almost every novel site, with no need for retraining nor fine-tuning, and with no relevant performance offset in segmenting either internal or external sites.
Furthermore, it proves to be general and robust across sites against different population demographics, anatomical challenges, clinical conditions, and technical specifications (\textit{e.g.}, field strength, manufacturer).

As an open source tool, \texttt{LOD-Brain} can be used off-the-shelf on unseen scans from novel sites.
Segmentation masks are returned very quickly (few seconds on a GPU) thanks to a reduced number of model parameters ($300K$), if compared to other state-of-the-art solutions.
To maximise research reproducibility and state-of-the-art comparisons, we adopt for testing the MICCAI anatomical structure labels proposed in \cite{mendrikMRBrainSChallengeOnline2015a}, using FreeSurfer \cite{fischlFreeSurfer2012a} segmentation masks as silver ground-truth (\textit{i.e.}, a ground-truth  with errors).
However, as we release both the model and the code at the \href{https://rocknroll87q.github.io/LOD-Brain/}{project website}, \texttt{LOD-Brain} can be retrained from scratch to deal with any set of structures and labels obtained by any manual or automatic software. 
A working demo is also available \href{https://rocknroll87q.github.io/LOD-Brain/}{here}. 
\section{Related work}
\label{sec:soa}

Atlas- or multi-atlas based methods, such as FreeSurfer \cite{fischlFreeSurfer2012a} or FLS~\cite{jenkinson2012fsl}, are still largely adopted for brain MRI segmentation~\cite{cabezasReviewAtlasbasedSegmentation2011}. 
Despite the needed registration procedure usually provides a good alignment between volumes, it requires hours of processing for each scan \cite{kleinMindbogglingMorphometryHuman2017}, thus imposing barriers to groups with limited computational capabilities in case of large-scale studies \cite{bethlehem2022brain}.
Furthermore, atlas-based strategies are hardly effective on data with abnormalities, either in terms of anatomy or intensity distributions, requiring manual intervention for fixing automatic errors.

In recent years, deep learning (DL) techniques deeply impacted medical imaging \cite{litjensSurveyDeepLearning2017} and image segmentation tools \cite{IJK21}.
Regarding the brain, first DL-based methods were limited in handling the 3D nature of MRI data, as they processed single 2D slices only. 
QuickNAT ~\cite{royQuickNATFullyConvolutional2019} tries to overcome the drawbacks imposed by 2D segmentation by aggregating the predictions of three different 2D slice-based encoder-decoder models, one per canonical slicing plane (longitudinal, sagittal, and  coronal), and combining the three results for obtaining the segmentation.
FastSurferCNN \cite{FastSurfer-HCE20} applies the same 2D approach training the network on a sequence of 2D neighbouring slices, instead of a single slice.
To reduce the loss of 3D context and minimise inter-slice artefacts, methods processing 3D-patches and aggregating the resulting sub-volumes are proposed in \cite{ dolzHyperDenseNetHyperdenselyConnected2018c, wachingerDeepNATDeepConvolutional2018}.
However, all these tools exploit only local 3D spatial information, while global spatial clues, such as the absolute and relative positions of different brain structures, are disregarded, hindering any possible learning of anatomical priors.
Other ensemble approaches based on multiple CNNs processing different overlapping brain sub-volumes, such as AssemblyNet \cite{coupe2020assemblynet} or SLANT \cite{huo20193D}, achieve whole brain segmentation, at the cost of an explosion of parameter cardinality. 
To avoid these drawbacks typical of the tiling process on 2D or 3D patches \cite{ReinaFrontiers2020}, CEREBRUM tools represent a fully 3D solution to brain MRI segmentation for 3T \cite{bontempi2020cerebrum}, and 7T scans \cite{svanera2021cerebrum}. 
However, similarly to DL methods which are trained on single-site MRIs, they also do not perform well on volumes from unseen sites, as they require training from scratch, or fine-tuning for each new target distribution \cite{svanera2021cerebrum}.

Data harmonisation strategies, when oriented to an explicit removal of site-related effects in multi-site data \cite{pomponio2020harmonization}, constitute a valid strategy to partially alleviate the unwanted performance drop due to the scanner effect.
To mitigate inter-site differences, Beer et al. propose in \cite{beer2020longitudinal} a longitudinal version of the ComBat method: an empirical Bayesian approach which applies a multivariate linear mixed-effects regression to account for both the biological variables and the scanner.
The model is able to adjust for additive and multiplicative effects by calculating a site-specific scaling factor.
A joint normalising function across multiple datasets is instead learnt by Delisle et al. in \cite{delisle2021realistic} by means of two fully-convolutional 3D CNNs: the first normalises image intensities across multiple datasets, while the second optimises images for a downstream segmentation task. 
Despite harmonisation algorithms mitigate scanner-specific effects, they not always preserve the inter-subject biological variability from each site, and are sometimes sensitive to changes in pre-processing steps \cite{cetin2020exploring}.
 
Closely related to harmonisation, domain adaptation methods try to adapt the segmentation networks trained on a source domain to produce correct outputs also on samples from a target domain.
As an example, DeepHarmony \cite{dewey2019deepharmony} exploits a fully-convolutional CNN architecture to map brain scans of a subject from one source acquisition protocol to a target one.
However, DeepHarmony cannot be extended to more than two sites since it relies on learning a protocol-to-protocol mapping.

SynthSeg \cite{billot2021synthseg} is an effective adaptation method which, starting from a full domain randomisation of the training set, segment brain MRI scans of any contrast and resolution, without retraining nor fine-tuning. 
As traditional data augmentation has limited ability to emulate real variations, SynthSeg is trained with synthetic scans obtained by leveraging a generative model with fully randomised parameters (intensity, shape, etc.). 
Despite its high accuracy, peculiar scanner artefacts and the absence of alignment parameters in the image header determine the presence of errors in the segmentation.

Far from applying full domain randomisation, Zhao et al. \cite{zhao2019data} propose an alternative but still aggressive augmentation solution. 
This approach first learns independent spatial and appearance transform models to capture the variations in a dataset of brain scans. 
Then, it uses these transform models to synthesise a dataset of labelled examples starting from only a single selected scan. 
The synthesised dataset is eventually used to train a supervised network, which significantly improves over previous methods for one-shot biomedical image segmentation, but with unclear outcomes in the presence of larger labelled training sets.
Other synthetic approaches adopt generative adversarial networks (GAN) to create synthetic abnormal MRI images with brain tumours, so as to improve tumour and brain segmentation~\cite{shin2018medical}.
Despite synthetic methods increase generalisation, aggressive augmentations do not always represent a solution for coping with distinct scanners and protocols, especially if they do not increase model performance.

The first multi-site attempt of gaining a model which is robust to the scanner effect is described in \cite{liu2020MS-Net-multisite} in the domain of prostate segmentation.
Authors first perform feature normalization for each site separately, and then extract more generalizable representations from multi-site data by a novel learning paradigm.
Other works that adopt deep learning techniques to cope with the multi-site variability can be found in \cite{rundo2019use} again for prostate segmentation, and in \cite{dou2020unpaired} for multi-organ segmentation from unpaired CT and MRI.
However, most of these approaches confirm to perform well on internal subjects, whereas require additional external images for the adaptation step (\textit{e.g.,} see \cite{karani2018lifelong}) to adequately cope with testing data obtained using different imaging protocols or scanners.
The usage of preprocessing steps confirms that efficiently handling multi-site data is still an open challenge and how the development of models able to jointly handle structure segmentation and site adaptation is highly needed.
Learning directly from out-of-the-scanner MRI brain volumes (\textit{i.e.}, with no atlas-based pre-alignment) from multiple-sites, with no fine-tuning nor adaptation steps, is an option that has remained unexplored until now, despite the recent availability of a large amount of brain open data repositories.

\section{Brain MRI multi-site data}
\label{sec:data}

To address the huge brain MRI variability in intensity statistics and scanning artefacts, we collect almost 27,000 brain T1-weighted volumes of both healthy and clinical subjects, mainly scanned with \texttt{mprage/mp2rage} sequences, and released in 79 databases covering approximately 160 world sites\footnote{From open repositories, it is not always possible to retrieve the number and model of the scanners employed in acquisitions, nor the number of unique participants.
This means that a database could contain more than one site.
Henceforth, we try to distinguish between \textit{dataset} and \textit{site} wherever possible.}. 
We first aggregated data from well-known open repositories, such as \href{http://www.humanconnectomeproject.org/}{HCP}, \href{https://abcdstudy.org/}{ABCD}, \href{https://www.oasis-brains.org/}{OASIS}, and datasets contained in the \href{https://fcon_1000.projects.nitrc.org/}{INDI project} including  \href{http://fcon_1000.projects.nitrc.org/indi/enhanced/}{NKI-RS}, \href{https://brain-development.org/ixi-dataset/}{IXI},  \href{http://fcon_1000.projects.nitrc.org/indi/abide/}{ABIDE}, and \href{https://fcon_1000.projects.nitrc.org/indi/adhd200/}{ADHD}.
Then we added datasets from open platforms as \href{https://openneuro.org}{OpenNeuro}, \href{https://osf.io/}{OSF}, \href{https://www.neugrid2.eu/index.php/data-portfolio/}{neuGRID2} and \href{https://nda.nih.gov/}{NIMH}, avoiding paid repositories such as \href{https://www.ukbiobank.ac.uk/}{UKBiobank}.
Other public datasets included are \href{https://mindboggle.info/data.html}{Mindboggle101}, \href{https://nilab-uva.github.io/AOMIC.github.io/}{AOMIC}, and \href{https://www.nitrc.org/projects/ibsr}{IBSR}.
Apart from Glasgow data\footnote {Maintained by authors, with pending ethics permissions for sharing.}, all repositories are available without fees, to maximise the reproducibility of this work.
A full data table is provided on the \href{https://rocknroll87q.github.io/LOD-Brain/dataset}{project website}.

\begin{figure*}
    \centering
    \includegraphics[width=1\textwidth]{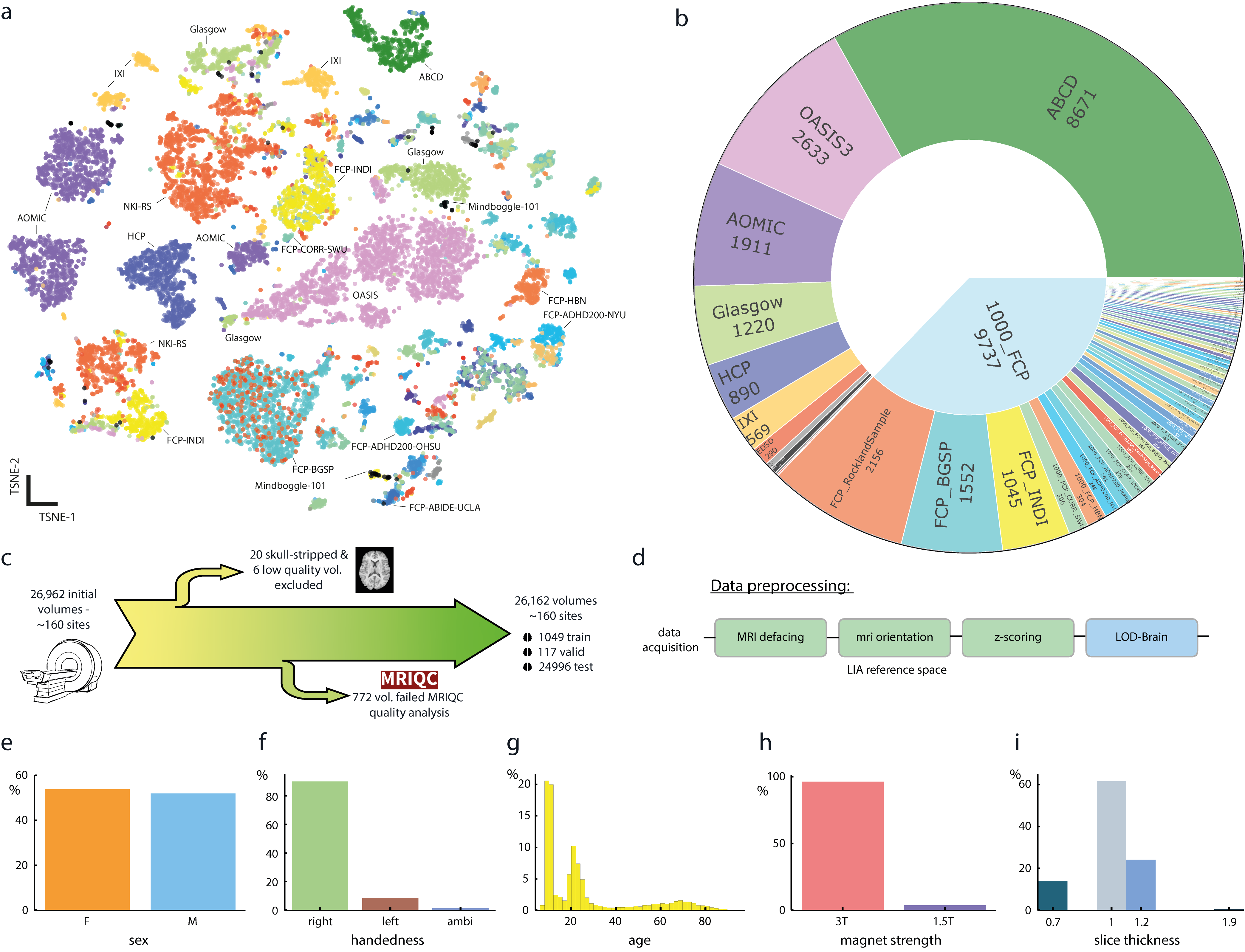}
    \caption{
    Multi-site dataset: we collect and analyse with MRIQC \cite{esteban2017mriqc} almost 27,000 volumes originating from around 160 different sites (26,169 volumes after the quality check). (a) A visualisation by t-SNE \cite{tsne} of the 68 MRIQC features (different colour for each dataset). Note that one dataset (\textit{e.g.}, \href{https://brain-development.org/ixi-dataset/}{IXI} in yellow colour) may contain volumes from more than one site or acquired with different scanners, and thus separate in clusters in the t-SNE space. (b) Dataset cardinalities. (c) Details on data quality assessment and (d) pre-processing. From (e) to (i), different demographic features and scanner properties are reported.}
    \label{fig:database}
\end{figure*}

In Figure~\ref{fig:database}, we present the composition of the dataset, its cardinalities and features, the quality assessment process done by MRIQC \cite{esteban2017mriqc}, and details on training and testing splits.
The 26,169 volumes that passed the MRIQC quality control analysis undergo defacing first, and then simple pre-processing steps before neural network feeding, including FreeSurfer's  \texttt{mri\_convert} to reorient volumes to LIA (left, inferior, anterior) reference space, padding to $256^3$, and z-scoring.

\subsection{Data split and labelling}

Out of the 79 datasets, 70 are considered as internal (INT), while 7 are left out for testing only (EXT).
The 2 remaining sets are used for specific analyses: SIMON~\cite{duchesne2019canadian} contains scans from a single healthy individual who participated in a multi-centre study; the last is a dataset with five patients with only one brain hemisphere from Kliemann et. al ~\cite{kliemann2019intrinsic}.
As validated in Section~\ref{ssec:exp-cardinalities}, the model used for testing is trained on a randomised selection of 1,049 volumes from internal data (15 volumes for each dataset, except one contributing with 14 volumes as it does not have enough data).
This allows to obtain a balanced training set in terms of dataset representativeness and an appropriate total number of training volumes for the learning task. 
The 77 datasets used for testing (70 INT and 7 EXT) include a total of 24,996 volumes (15,841 INT and 9,155 EXT).
Since only $10\%$ of the datasets include more than $80\%$ of the testing volumes, we select up to 200 volumes per dataset to avoid biases and guarantee balanced results, ending up with a total of 5,956 testing volumes (5,360 INT and 596 EXT).
The validation set, used for hyperparameter selection, includes 117 volumes from 72 datasets (91 INT and 26 EXT).

As no manual segmentations (gold standard) are available for most volumes, training adopts a weakly supervised learning strategy, exploiting segmentation labels obtained by FreeSurfer \cite{fischlFreeSurfer2012a} as a silver standard ground-truth (GT), similarly to what proposed in \cite{bontempi2020cerebrum}.
The only dataset with semi-manual labels \textit{i.e.}, \href{https://mindboggle.info/data.html}{MindBoggle} (FreeSurfer plus manual corrections), is exploited in validation and testing.
The manual segmentations provided for \href{https://www.nitrc.org/projects/ibsr}{IBSR} and \href{http://www.neuromorphometrics.com/2012_MICCAI_Challenge_Data.html}{MALC2012} were discarded and replaced with FreeSurfer outputs, because of their low quality.
As for the quality of the FreeSurfer's GT masks, these present high variability.
In particular, out of the seven external datasets (testing only), four present an acceptable GT (covering a total of 32 sites), while the other three show low quality GT segmentations as they include clinical scans.
Low quality GT masks are usually produced from low quality T1w volumes; while they are not used for training, since we do not want to compromise the model learning ability, they are still used for testing to explore model capabilities and limitations.

The labelling strategy follows the 7 classes adopted by MRBrainS challenge \cite{mendrikMRBrainSChallengeOnline2015a}: grey matter, white matter, cerebrospinal fluid, ventricles, cerebellum, brainstem, and basal ganglia.
Such labeling maximises the possibility of comparison with other state-of-the-art methods, and covers most of clinical and research studies and applications.
However, there are no limitations in selecting different brain structures and related labels for retraining \texttt{LOD-Brain}.

\section{Methods}
\label{sec:methods}

\subsection{Architecture: a 3D level-of-detail network}

\texttt{LOD-Brain} is a progressive level-of-detail 3D network designed for brain MRI segmentation.
As shown in the general scheme in Fig.~\ref{fig:framework}, each level of \texttt{LOD-Brain} is a U-net~\cite{cicek3DUNetLearning2016b} which processes the input MRI volume (of initial dimensions $256^3$) at a different scale obtained by successively down-sampling the volume by a factor $d$ along each coordinate axis.
The lowest network level, $LOD_{L}$, is in charge of learning a robust \textit{anatomical prior}.
Since down-sampling input volumes removes high frequency details and smoothes individual differences, $LOD_{L}$ learns a coarse representation of brain structures, and their mutual locations, which is less dependent on the scan site.
Training happens in a bottom-up approach: after convergence, $LOD_{L}$ is frozen, and inter-level connections ensure that the 3D spatial context learnt at the lower level is embedded and propagated to $LOD_{L-1}$ and, from there, to higher levels of the architecture.
The process is iteratively repeated through superior levels until the upper one \textit{i.e.}, $LOD_1$, which processes the input data at the fullest scale, refining the segmentation masks at the finest detail and accounting for site-specific intensity distributions.

The loss $\mathcal{L}$ adopted by \texttt{LOD-Brain} is a mixed per-channel dice function $\mathcal{L}_{dice}$ and cross entropy loss $\mathcal{L}_{CE}$: 
$$\mathcal {L} = -\lambda \mathcal {L_{CE}} -(1-\lambda) \mathcal {L}_{dice}$$
with $\lambda$ balancing the two components.  
In particular, $\mathcal{L}_{CE}$ is:
$$\mathcal {L}_{CE} = \sum_{i=1}^{C} \sum _{k=1}^{V} y_{k,i} log(F_{k,i})
$$
where $V$ and $C$ are the set of voxels and classes,
respectively, $y$ is the GT mask, and $F$ is the output.
Conversely, $\mathcal {L}_{dice}$ is:
$$
\mathcal {L}_{dice} = 1 - \frac{2}{C} \sum_{i=1}^{C} \frac{\sum _{k=1}^{V}  y_{k,i} F_{k,i}}{\sum _{k=1}^{V} y_{k,i}^2 + F_{k,i}^2}
$$
Hyperparameter selection, network design, and the choice of parameters $L$, $\lambda$, $d$, etc. is described in Section~\ref{sssec:implementation}.

\begin{figure}[h]
    \centering
    \includegraphics[width=0.95\columnwidth]{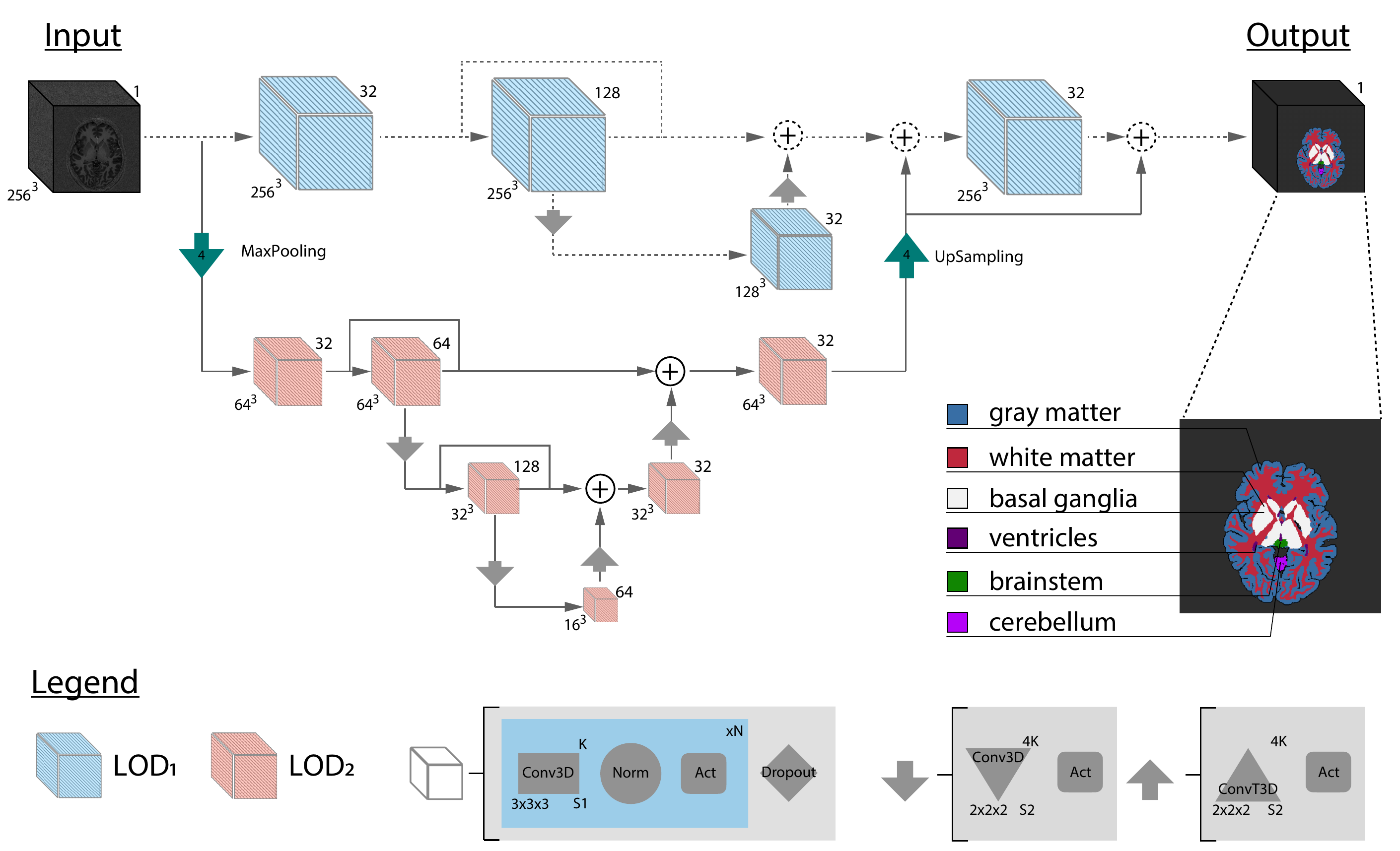}
    \caption{\texttt{LOD-Brain} architecture selected for the experiments on the brain MRI segmentation task ($L=2$, $d=4$).
    }
    \label{fig:model}
\end{figure}

The architecture which emerges as the best performing one during the experiments is presented in Fig.~\ref{fig:model}.

The network is made up of three basic 3D convolutional blocks.
The first addresses feature learning: it is composed of a $3\times3\times3$ convolution layer followed by normalisation and non-linear activation, all repeated multiple times, ending with a dropout layer.
The other two blocks perform down-sampling and up-sampling, with strided convolution and transposed convolutions, respectively, both followed by non-linear activations.
These layers allow the network to learn optimal up/down-sampling strategies and process different extracted feature hierarchies.
Moreover, skip connections and inter-level connections are implemented along with summation nodes, as it was proven to have a better trade-off between segmentation accuracy and parameter count compared to concatenation
\cite{milletariVNetFullyConvolutional2016a}.

\subsection{Data augmentation}
\label{ssec:data_augm}

Instead of performing a pre-selected set of common data augmentations, we perform an ad-hoc procedure to verify the usefulness of augmentations in advance.
In the first step, we create a pool of realistic transformations belonging to three categories: geometrical transformations, noise distortions, and artefact introduction.
In the first category, in addition to classical operations such as, flip, rotation, and translation, we also introduce grid distortion.
The second category accounts for a comprehensive set of noises: salt and pepper, Gaussian, Gamma, and contrast noise.
The last transformation family focuses on mimicking MRI artefacts like ghosting and MR field inhomogeneity, as described in~\cite{svanera2021cerebrum}.
In the second step, we test which transformation is beneficial to increase the model performance.
Validation is done by applying each transformation to the validation set volumes (by increasing transformation parameters), and then computing the performance of a model trained without any data augmentation.
If the model is already robust to a specific transformation (\textit{i.e.}, there is no performance gap in testing a volume with and without transformation), this is no further considered.
Otherwise, in those situations in which the training set is not rich enough (\textit{i.e.}, whenever transforming the input data introduces a performance drop), such transformation is considered suitable for augmentation, since it can introduce a realistic alteration to input volumes that the model is not able to handle yet.
Table \ref{tab:augmentation} reports details on selected augmentations only, showing probabilities of application and parameters justified by the experiments detailed in Section~\ref{subsub:augmentation}.

\begin{table}[h]
\caption{Details on selected augmentation methods.}
\scriptsize
\centering
\label{tab:augmentation}
\begin{tabular}{@{}llcl@{}}
\toprule
Group                        & Augmentation          & Prob. & Parameters                   \\ \midrule
\multirow{2}{*}{Geometrical} & Flip         & 1/2         & Sagittal plane only                            \\
                             & Grid distortion       & 1           & Steps: 5; Distortion: .1     \\ \midrule
\multirow{6}{*}{Noise}       & Salt and pepper       & 1/6           & Amount: 0.01; Salt: 0.2      \\
                             & Gaussian              & 1/6           & Amount: 0.2                  \\
                             & Gamma                 & 1/6           & Clip: 0.025                  \\
                             & Contrast              & 1/6           & Alpha: 0.5-3.0               \\
                             & Blur                  & 1/6           & Limit: 3                     \\
                             & Downscale             & 1/6           & Scale: 0.25-0.75             \\ \midrule
\multirow{2}{*}{Artefacts}   & Ghosting              & 1/2           & Max rep.: 4                 \\
                             & Inhomogeity           & 1/2           &  See \cite{svanera2021cerebrum} for details                            \\
                              \bottomrule 
\end{tabular}
\end{table}

\section{Results and Discussion}
\label{sec:results}

The experimental assessment of our multisite-based model is structured as follows.
The first set of experiments aims to justify the choice of the adopted model. 
Next, we test the robustness and generality of \texttt{LOD-Brain} on different types of data (internal and external datasets, and data with marked anatomical variations), and the invariance of the model against different types of bias, including scanner vendors and models.
Eventually, we quantitatively and qualitatively compare our method against state-of-the-art.
Unless differently stated, results are computed using Dice coefficient as performance metric, and FreeSurfer segmentation as silver ground-truth. 

\subsection{Model training and hyperparameter selection}

\subsubsection{Multi-site learning}
\label{ssec:exp-cardinalities}
Given the richness of the aggregated data, we first want to find suitable dimensions for the training set. 
This requires answering to the following questions: in order to reach good generalisation capability (\textit{i.e.}, similar performance on both INT and EXT testing data), how many volumes per site should be considered in the training set? 
And, how many different datasets should be included?   

\begin{figure}[h]
     \centering
     \begin{subfigure}[b]{0.49\columnwidth}
         \centering
         \includegraphics[width=\textwidth]{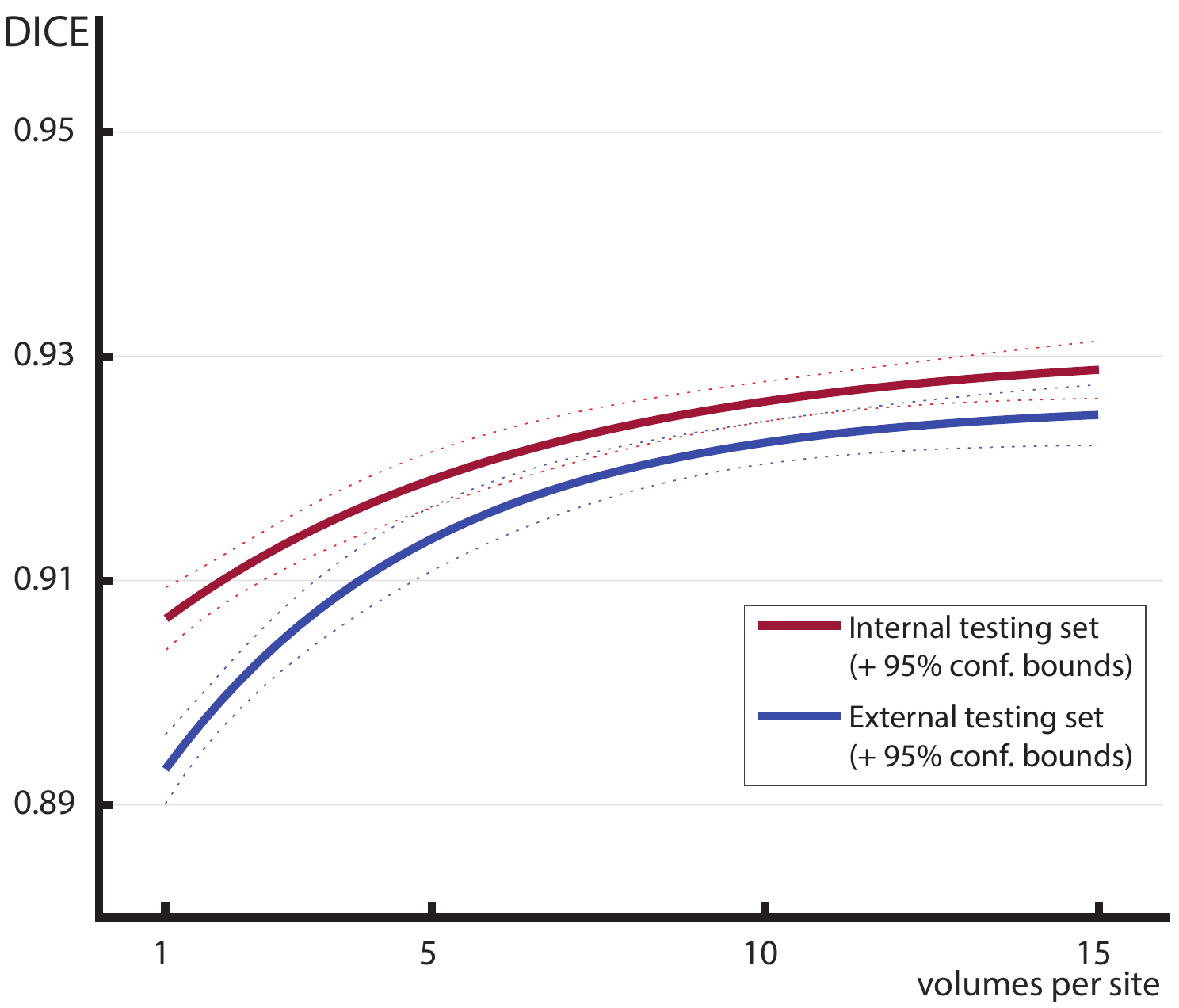}
         \caption{How many volumes?}
         \label{fig:cardinality_multisite_how_many_samples}
     \end{subfigure}
     \begin{subfigure}[b]{0.49\columnwidth}
         \centering
         \includegraphics[width=\textwidth]{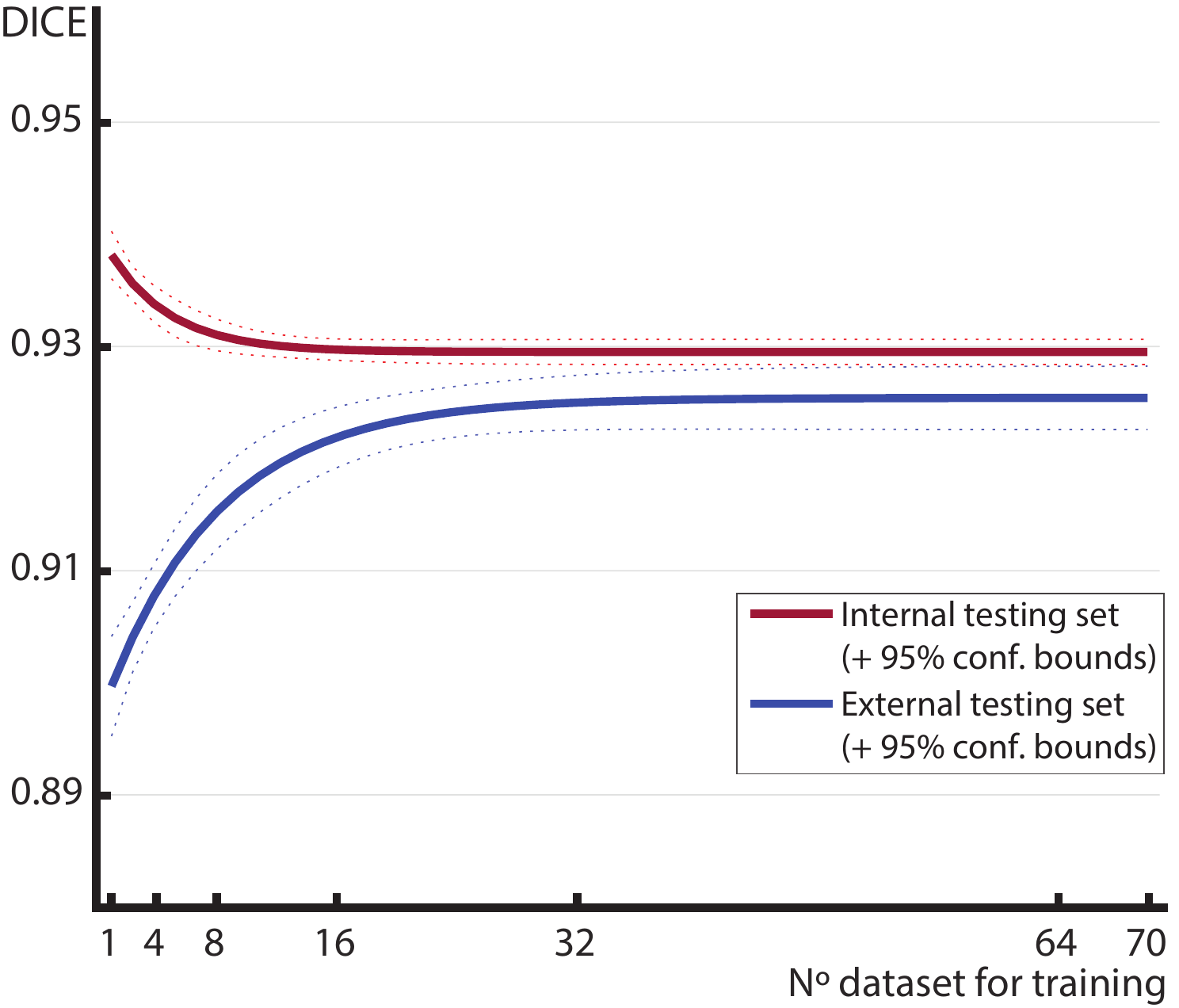}
         \caption{How many datasets?}
         \label{fig:cardinality_multisite_how_many_site}
     \end{subfigure}
        \caption{
            (a) Using 70 available training datasets, we trained 4 models with $[1, 5, 10, 15]$ volumes per dataset. The model is tested on INT (red) and EXT data (blue).
            (b) Using 15 volumes per dataset, we train models with an increasing number of sites $[1, 4, 8, 16, 32, 64, 70]$. Testing is done on INT and EXT data (red and blue respectively). Both graphs fit exponential curves.
            }
        \label{fig:cardinality_multisite}
\end{figure}

In Fig.~\ref{fig:cardinality_multisite_how_many_samples}, we show the performance of models trained with 1, 5, 10, or 15 volumes per dataset, considering all $70$ available training datasets (\textit{i.e.}, number of training volumes = $70, 350, 700, 1049$).
Internal testing (INT) is performed on 110 unseen volumes equally distributed among the same datasets used for training at the considered step, while external testing includes again $110$ volumes from $4$ left out datasets (\href{https://abcdstudy.org/}{ABCD},
\href{http://www.neuromorphometrics.com/2012_MICCAI_Challenge_Data.html}{MALC2012}, 
\href{https://fcon_1000.projects.nitrc.org/indi/pro/eNKI_RS_TRT/FrontPage.html}{1000\_FCP\_CORR\_NKI\_TRT}, and \href{https://mindboggle.info/data.html}{MindBoggle101}).
As expected, using more training volumes per dataset enhances the segmentation accuracy for both INT and EXT testing data.
Since we reach a plateau of performance between 10 and 15 volumes per dataset, and to avoid the introduction of dataset-related biases, we decide to use 15 volumes per dataset, which is the maximum possible for maintaining balance across datasets. 

In Fig.~\ref{fig:cardinality_multisite_how_many_site}, we evaluate the accuracy of \texttt{LOD-Brain} as a function of the number of datasets included in training.
Testing volumes, both INT and EXT, are the same used in Fig.~\ref{fig:cardinality_multisite_how_many_samples} to allow comparisons.
For each value $i\in[1,4,8,16,32,64,70]$, we retrain \texttt{LOD-Brain} with $1,049$ volumes selected from $i$ datasets randomly chosen from those with enough samples.
As shown in Fig.~\ref{fig:cardinality_multisite_how_many_site}, as long as the number of sites increases, the gap of performance between internal and external testing data progressively decreases, until it fades.
Therefore, unless otherwise specified, we set to 70 the number of datasets used to train \texttt{LOD-Brain}.

\subsubsection{Parameter selection}
\begin{figure}[htb]
    \centering
    \includegraphics[width=\columnwidth]{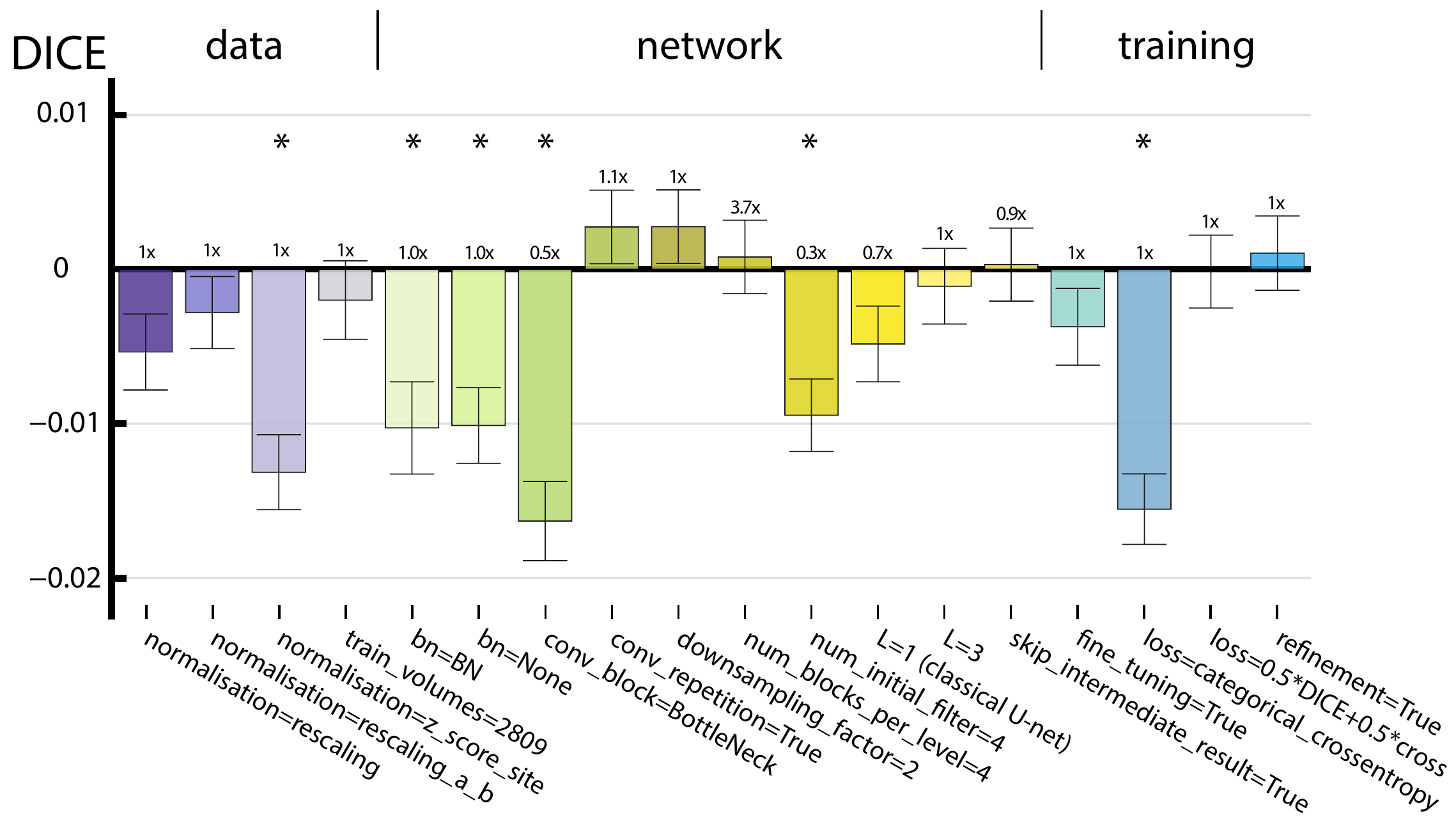}
    \caption{
     Ablation study. Performance of models trained with different architectural options are shown with respect to the best model (on the zero x-axis). Results (Dice coefficient differences) are computed on the validation set (those marked with * are statistically significant according to $t_{test}$ applying Bonferroni correction).
    }
    \label{fig:ablation_study}
\end{figure}

In Fig.~\ref{fig:ablation_study}, we present the most relevant results of the ablation study carried out to select model parameters.
This includes investigations regarding data processing, network architecture, and training.
All results are computed on the validation set by evaluating their statistical significance and, in case no significance is found, by preferring models with least parameters.

On data, we evaluate the most advantageous type of data normalisation, and we make an attempt to train with a larger training set (almost 3k samples). However, since this causes a unbalance in data, we observe a drop in performance.

Regarding the network, we test different design choices for its architecture \textit{e.g.}, the number of levels $L$, the convolutional block (plain or residual), layer normalisation (batch or group), etc.
As a result, the LOD network implemented for testing is configured on $L=2$ levels and a down-sampling factor of $d=4$, as shown in Fig.~\ref{fig:model}.
It is relevant to note that the two levels resulting from the ablation study recall somehow the effectiveness of the approach extensively used in the past for brain segmentation: atlas-based registration first, followed by voxel-level segmentation.
Similarly here, the coarser level learns a robust brain prior which replaces the registration step in identifying brain structure locations, while, the finest level, handles site-specific intensity distributions and artefacts.
The entire procedure also may resemble the steps of manual segmentation, in which the human expert first zooms out to identify major anatomical structures, and then zooms in refining structures until the task is complete at the finest level.

With respect to training choices, we compare, among others, different loss functions (best with $\lambda=0$ \textit{i.e.}, pure Dice loss) and evaluate as detrimental a refinement of the entire unfrozen network, thus confirming that the brain prior learnt at $LOD_2$ is robust, and that a joint fine-tuning with a higher level would negatively affect its site-independent brain representation.

\subsubsection{Data augmentation}
\label{subsub:augmentation}

\begin{figure}[htb]
    \centering
    \includegraphics[width=0.95\columnwidth]{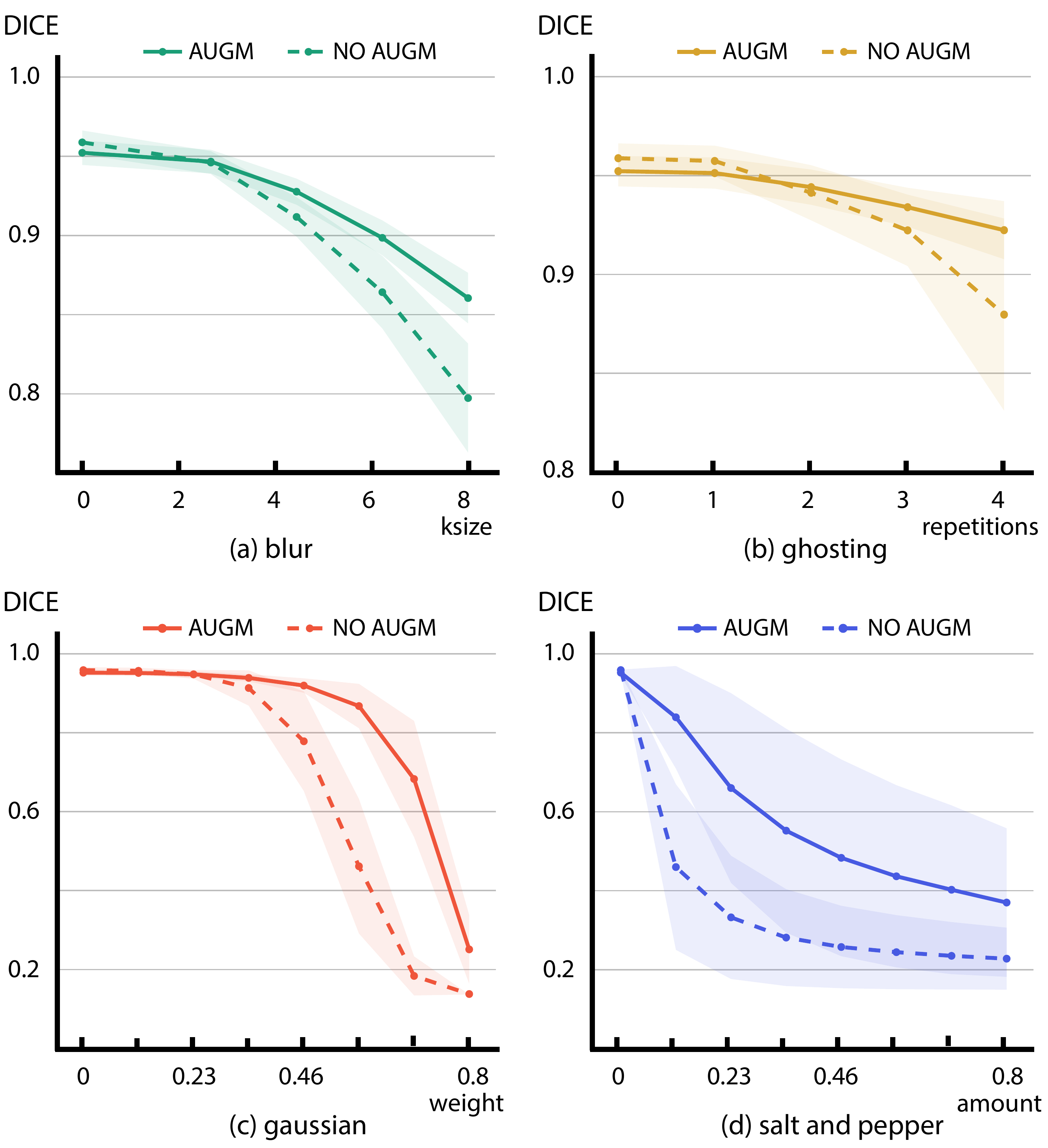}
    \caption{
    Data augmentation: performance of models trained with versus without augmentation for four transformations (\textit{i.e.}, blur, ghosting, gaussian, and salt and pepper noise). 
    }
    \label{fig:robustness_study}
\end{figure}

After selecting the useful transformations as in Section~\ref{ssec:data_augm}, we augment the validation set ($117$ volumes) and test the two models trained with and without augmentation.
Fig.~\ref{fig:robustness_study} reports the comparison as function of the augmentation parameters for four significant transformations.

\subsubsection{Implementation details}
\label{sssec:implementation}

Training optimization is done using Adam~\cite{kingma2014adam} and training lasts $50$ epochs for $LOD_2$ and $30$ for $LOD_1$, with an initial learning rate of $5e-4$, reduced by $1/4$ on plateau. 
As non-linear activation, \texttt{relu} is applied for both encoder and decoder.  
For better regularisation, each convolutional block performs group normalization, while the dropout rate is $0.05$.
Training lasts 3 days using a workstation with Nvidia$\textcopyright$ Quadro RTX 8000 GPUs and \href{https://www.wandb.com/}{\textit{Weights \& Biases}} for experiment tracking.

\subsection{Robustness and generalisability}
\label{ssec:robustness}

In this series of experiments, we test \texttt{LOD-Brain} on a variety of scenarios to assess its robustness and capabilities.

\subsubsection{Accuracy across datasets}

\begin{figure*}[htb]
     \centering
     \begin{subfigure}[b]{1.49\columnwidth}
         \centering
         \includegraphics[height=0.26\textheight]{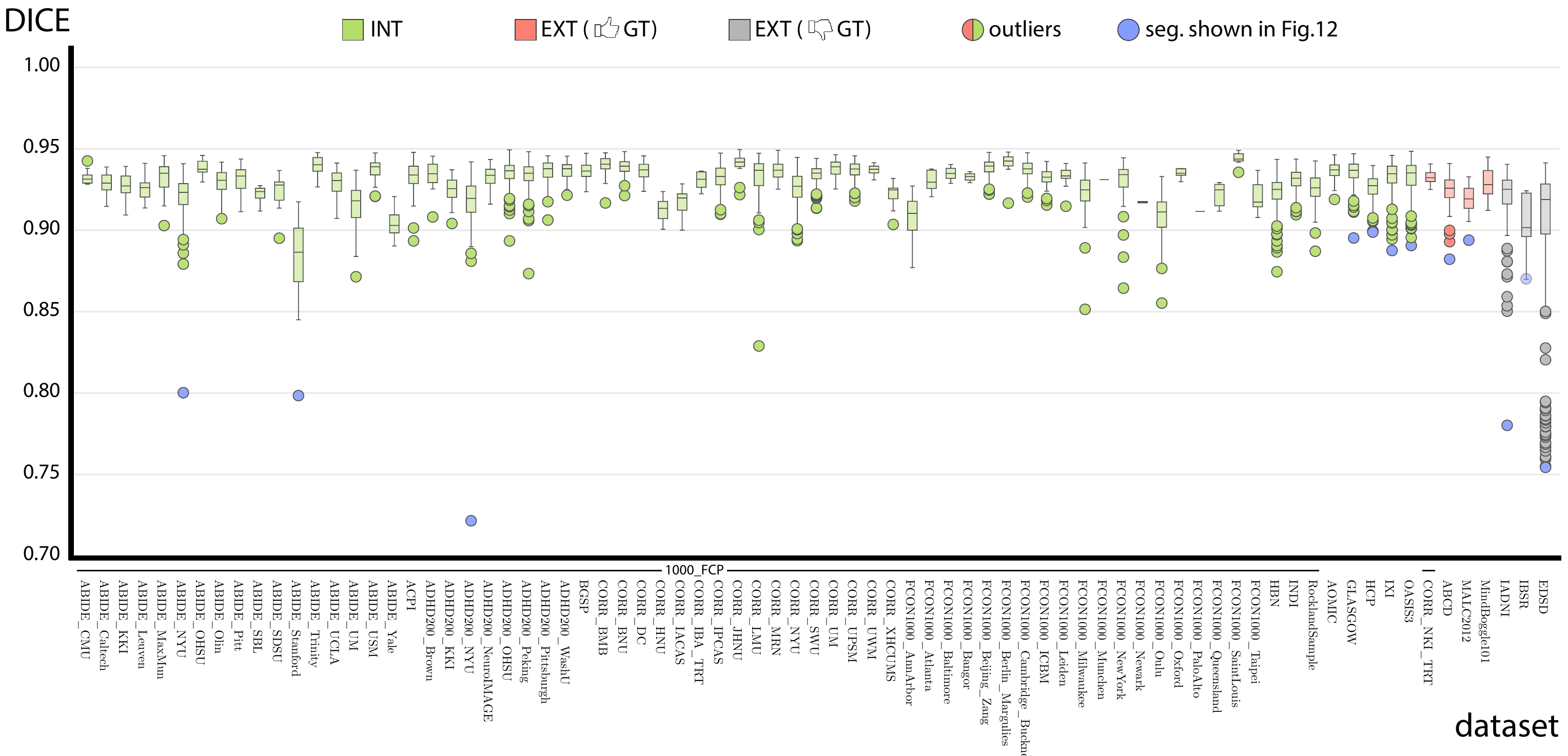}
         \caption{LOD-Brain results on all datasets}
         \label{fig:complete_testing_by_site}
     \end{subfigure}
     \begin{subfigure}[b]{0.5\columnwidth}
         \includegraphics[height=0.26\textheight]{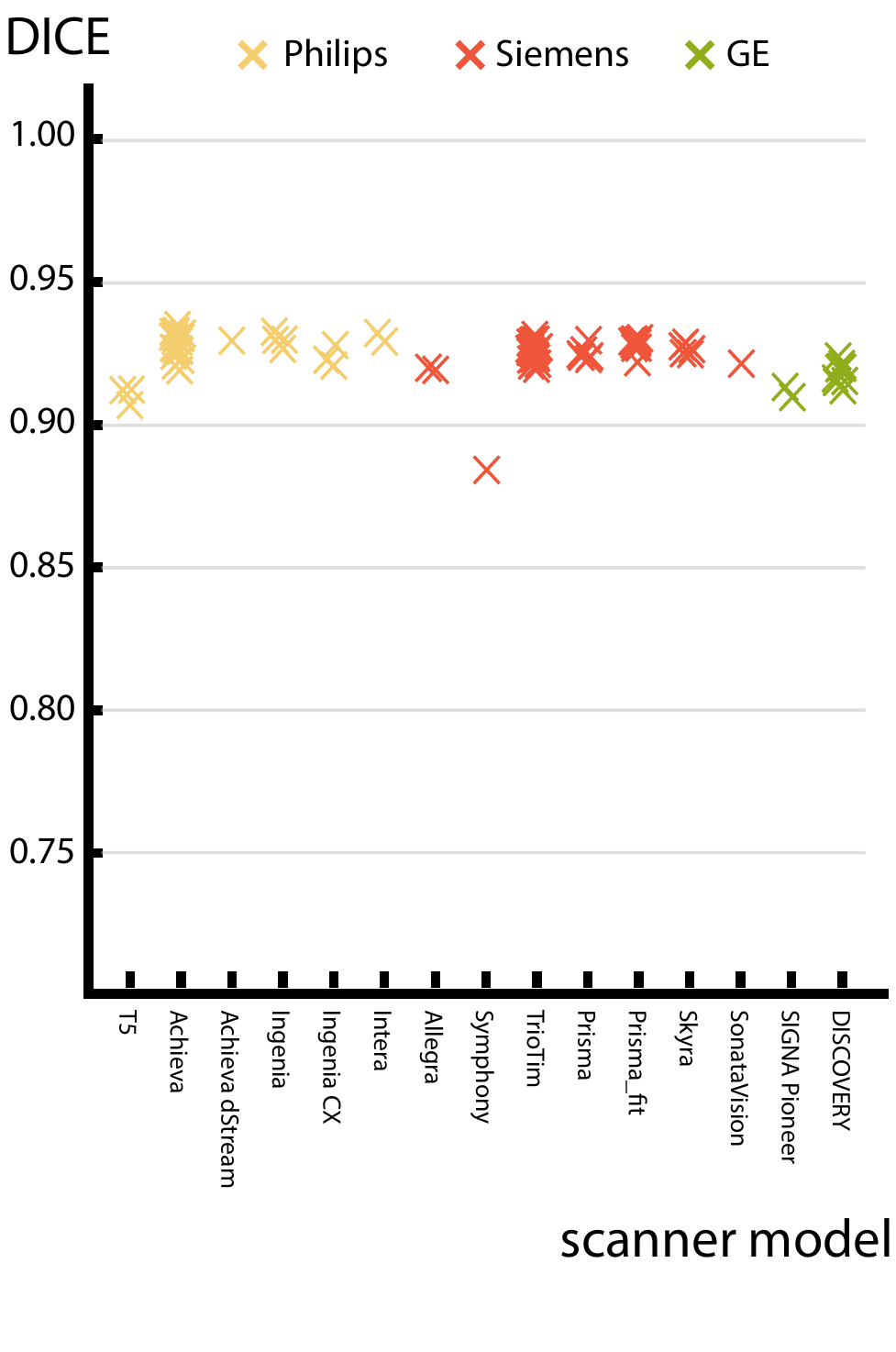}
         \centering
         \caption{Results on SIMON dataset}
         \label{fig:simon_results}
     \end{subfigure}
        \caption{
    Testing results. (a) Dice coefficient on $5,956$ testing volumes (77 datasets), displayed per dataset: INT (green), EXT with good GT (red), and with low-quality GT (grey). Segmentation masks of the worst numerical result (blue dots) are further displayed in Fig.~\ref{fig:error_analysis}.
    (b) Accuracy on SIMON (Single Individual volunteer for Multiple Observations across Networks) dataset (EXT) \cite{duchesne2019canadian}, comprising 94 volumes acquired with 15 different models of scanner by 3 major MR vendors (in different colours). 
            }
        \label{fig:results_testing}
\end{figure*}

Fig.~\ref{fig:complete_testing_by_site} reports segmentation performance for each of the 77 datasets used in this study. 
The overall accuracy (mean: $0.928$, std: $0.017$) proves the robustness of the method, showing similar results on both internal and external sites.
The performance obtained on low-quality GT datasets (in grey in Fig.~\ref{fig:complete_testing_by_site}) is justified by the presence of several scans with head movement artefacts due to participant populations (\textit{e.g.}, elderly people with dementia in \href{https://www.gaaindata.org/partner/EDSD}{EDSD} and children 7.5-12.9 y.o. in \href{https://fcon_1000.projects.nitrc.org/indi/abide/abide_I.html}{ABIDE\_Stanford}) which impair FreeSurfer segmentation.

\subsubsection{Multi-site versus single-site models}
\label{ssec:generalisation}

To validate the need for multi-site data, we compare the generalisation abilities of multi-site (MS) training with those of single-site (SS) models, by testing both (MS vs. SS) on the same internal (INT) and external data (EXT).
For enabling comparison, this experiment uses only the 4 single-site datasets with more than $1,049$ volumes (AOMIC: 1,911 volumes, Glasgow: 1,220, FCP\_BGSP: 1,552, FCP\_RocklandSample: 2,156).

\begin{figure}[htb]
    \centering
     \begin{subfigure}[b]{0.49\columnwidth}
         \centering
         \includegraphics[width=\textwidth]{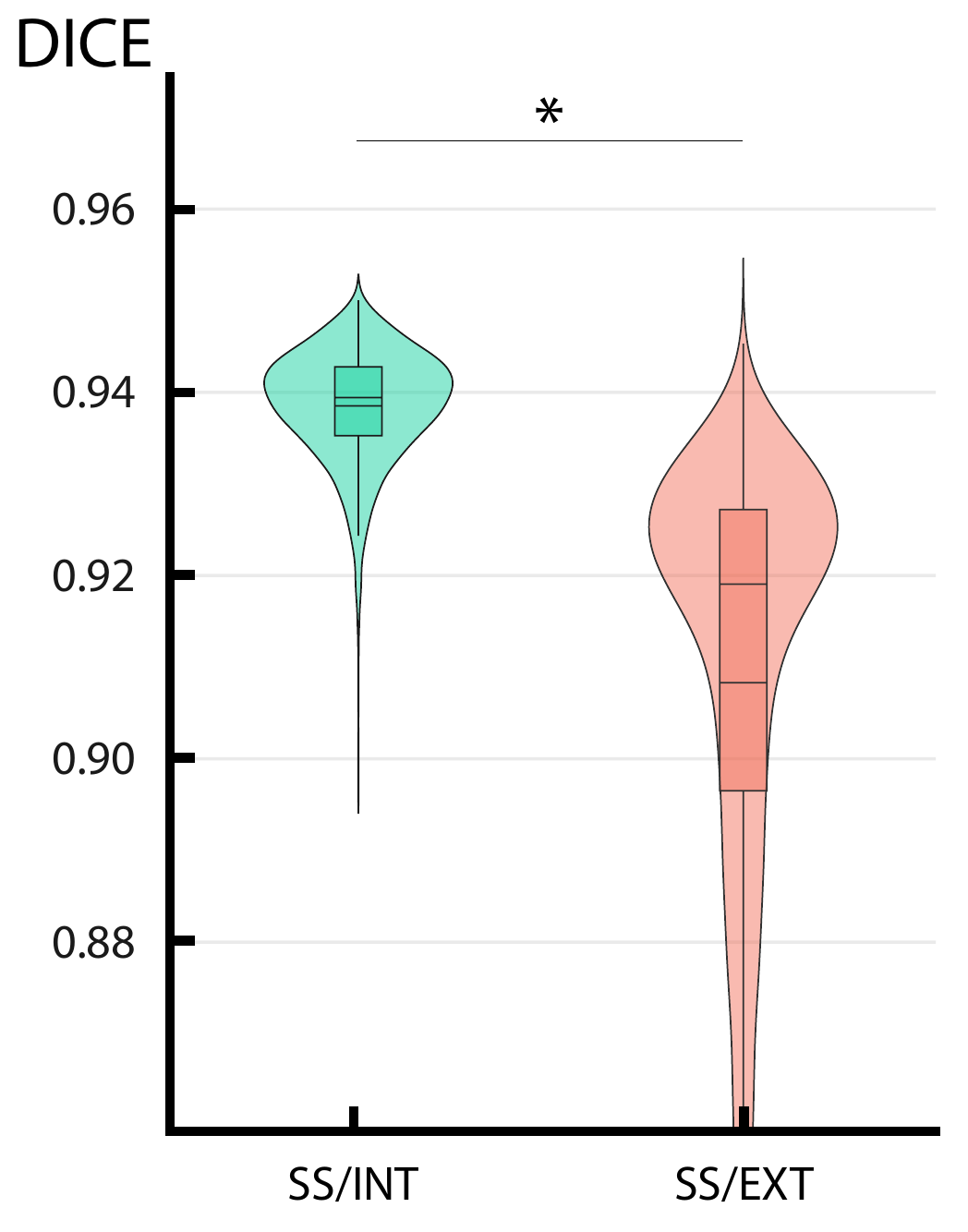}
         \caption{Single-site model}
         \label{fig:ss-vs-ms_SS}
     \end{subfigure}
     \begin{subfigure}[b]{0.49\columnwidth}
         \centering
         \includegraphics[width=\textwidth]{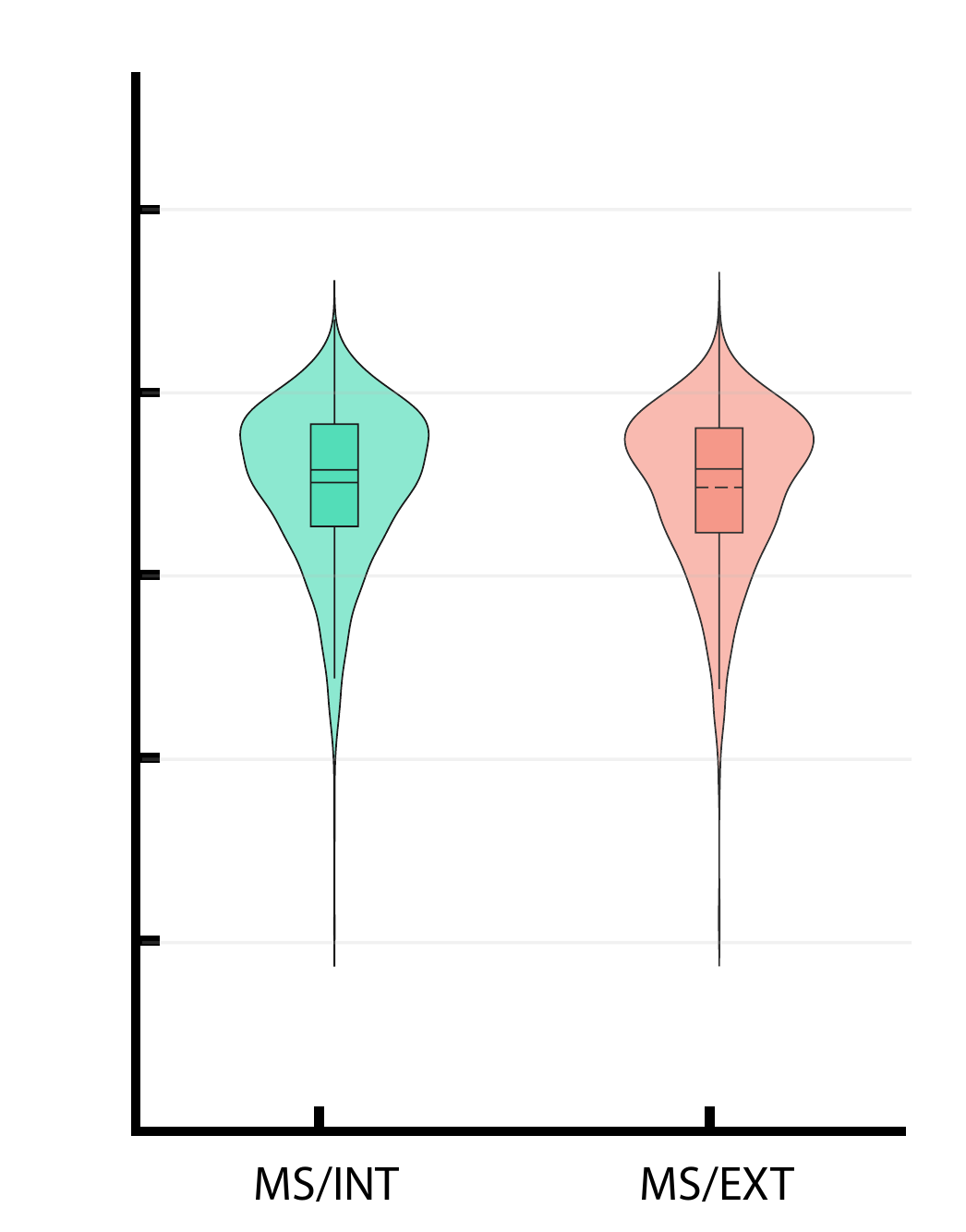}
         \caption{Multi-site model}
         \label{fig:ss-vs-ms_MS}
     \end{subfigure}
         \caption{
    (a) Performance obtained using single site (SS) models: in green, results on testing volumes from the same site (INT); in red, on testing volumes from different sites (EXT). 
    Performed $t_{test}$ shows a statistically significant difference (marked with *) between INT and EXT ($p_{value}< 1e^{-5}$).\\
    (b) Performance of multi-sites models (MS): in green, results on testing volumes from the same sites (INT); in red, on testing volumes from different sites (EXT) (no statistically significant difference between INT and EXT $p_{value}=0.11$).}
    \label{fig:ss-vs-ms}
\end{figure}

For each of the 4 datasets, we train a SS model with $1,049$ volumes, and we test its segmentation accuracy on both internal data (\textit{i.e.}, all remaining volumes from the same site) and external sites (\textit{i.e.}, all left-out volumes from the other 3 datasets).
A significant drop between INT and EXT performance, due to the scanner effect, is observed in Fig.~\ref{fig:ss-vs-ms_SS}.

As for multi-site training, we test our model trained with $1,049$ volumes from $70$ datasets on the same INT dataset used in the single-site case.
To test on external data, we train $4$ additional MS models on $69$ datasets, considering the left-out dataset (one among AOMIC, Glasgow, FCP\_BGSP, and FCP\_RocklandSample) as EXT data.
As both experiments in Fig.~\ref{fig:ss-vs-ms_SS} and Fig.~\ref{fig:ss-vs-ms_MS} use the same testing sets, we observe that models trained on multiple sites almost reaches on EXT data the same performance of SS models tested on unseen volumes from their training sites, while exhibiting far superior generalisation ability (\textit{i.e.}, a non-significant performance difference between INT and EXT data in Fig.~\ref{fig:ss-vs-ms_MS}).





%

\subsubsection{SIMON dataset}

As segmentation performance can vary for both scanner intensity distribution and variability in participants' anatomy, here we attempt to disentangle the two components.
We therefore test our model on a left-out dataset where, in the context of a multi-centre study \cite{duchesne2019canadian}, a single subject has been scanned many times and sites during his lifetime from 29 to 46 years old.
The dataset includes 73 sessions (94 volumes), 33 world locations, 15 different models of scanner, covering the 3 major MR vendors (GE, Philips, and Siemens). 
Fig.~\ref{fig:simon_results} results show an impressively high coherence among a large variety of scanner models by different vendors.

\subsubsection{Robustness to challenging anatomical variations}

\begin{figure}
     \centering
     \begin{subfigure}[b]{1.\columnwidth}
         \centering
         \includegraphics[width=\textwidth]{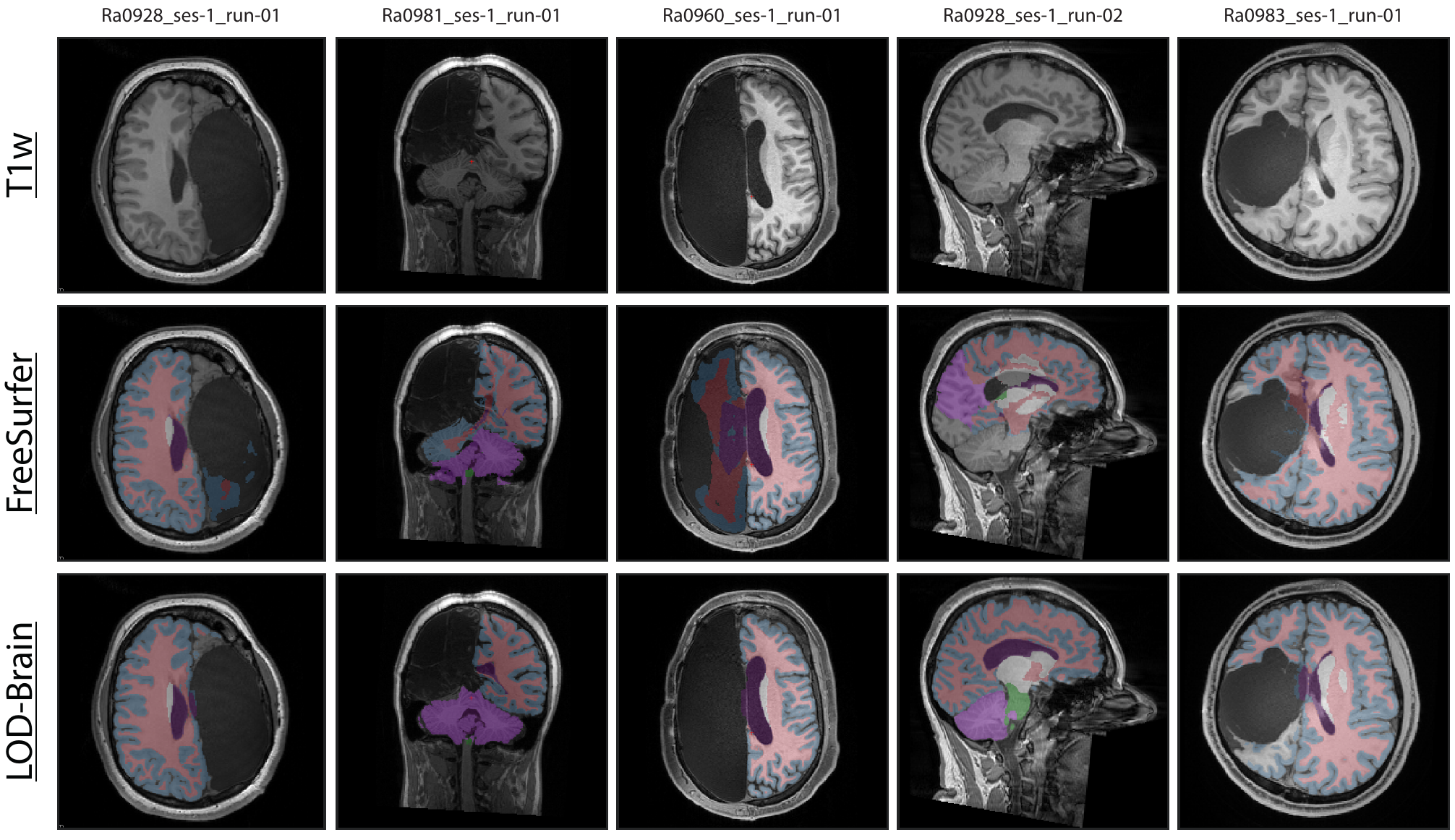}
         \caption{Results}
         \label{fig:half_brain_results-seg}
     \end{subfigure}
     \begin{subfigure}[b]{1.\columnwidth}
         \includegraphics[width=\textwidth]{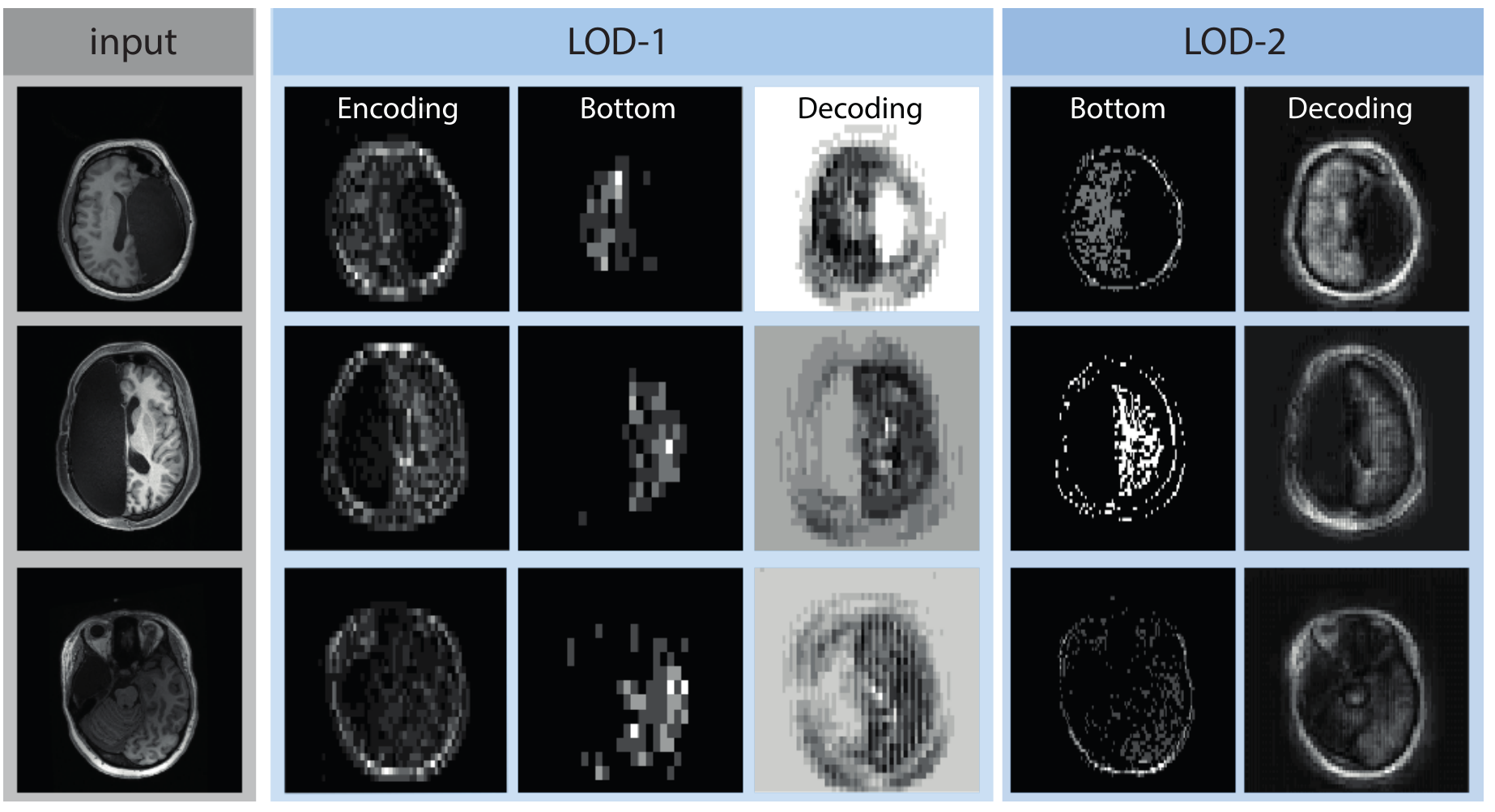}
         \centering
         \caption{Activation maps}
         \label{fig:half_brain_results-activ}
     \end{subfigure}
        \caption{
            Results on individuals who had undergone surgical removal of one hemisphere \cite{kliemann2019intrinsic}. (a) Inference for 5 subjects are shown for LOD-Brain and FreeSurfer. (b) Activation maps for 3 subjects at different LODs \textit{i.e.}, layers in the network.
            }
        \label{fig:half_brain_results}
\end{figure}

To test the robustness of the anatomical prior learnt, we test our model on a challenging scenario: a dataset with five individuals who had undergone surgical removal of one hemisphere \cite{kliemann2019intrinsic}.
In Fig.~\ref{fig:half_brain_results-seg}, we show the visual results obtained by FreeSurfer (second row), and by our method (third row).
While FreeSurfer - and in general atlas-based methods - fails to generalise to such severe anatomical singularities, often inferencing non-existing structures, \texttt{LOD-Brain} reliably segments such cases, proving a high level of robustness to anatomical variations.
In Fig.~\ref{fig:half_brain_results-activ}, we report different activation maps for three subjects of this dataset coming from different levels (i.e., layers) of the network. 
Skull stripping and cortex extraction is coarsely done already in $LOD_2$ (bottom layer) and then the information is propagated along the network upper levels. 
This result gives an intuition on how the coarse level acts as a prior, giving guidance to $LOD_1$ for finer segmentation.

\subsubsection{Invariance to bias}
To investigate the fairness of our segmentation model, we assess \texttt{LOD-Brain} for potential bias regarding demographic characteristics such as sex and age, or other technical characteristics of the scanner including scanner model, vendor, magnet strength, and slide thickness.
Despite the training data imbalance for some of these characteristics (see those in Figs.~\ref{fig:database} g/h/i), on the test-set of $5,956$ volumes we observe no salient differences in Dice performance between different groups.
Results are reported on the \href{https://rocknroll87q.github.io/LOD-Brain/results}{project website}.

\subsection{Method comparisons}
\label{ssec:soa-comparison}

\begin{figure}[htb]
    \centering
     \begin{subfigure}[b]{\columnwidth}
         \centering
         \includegraphics[width=\textwidth]{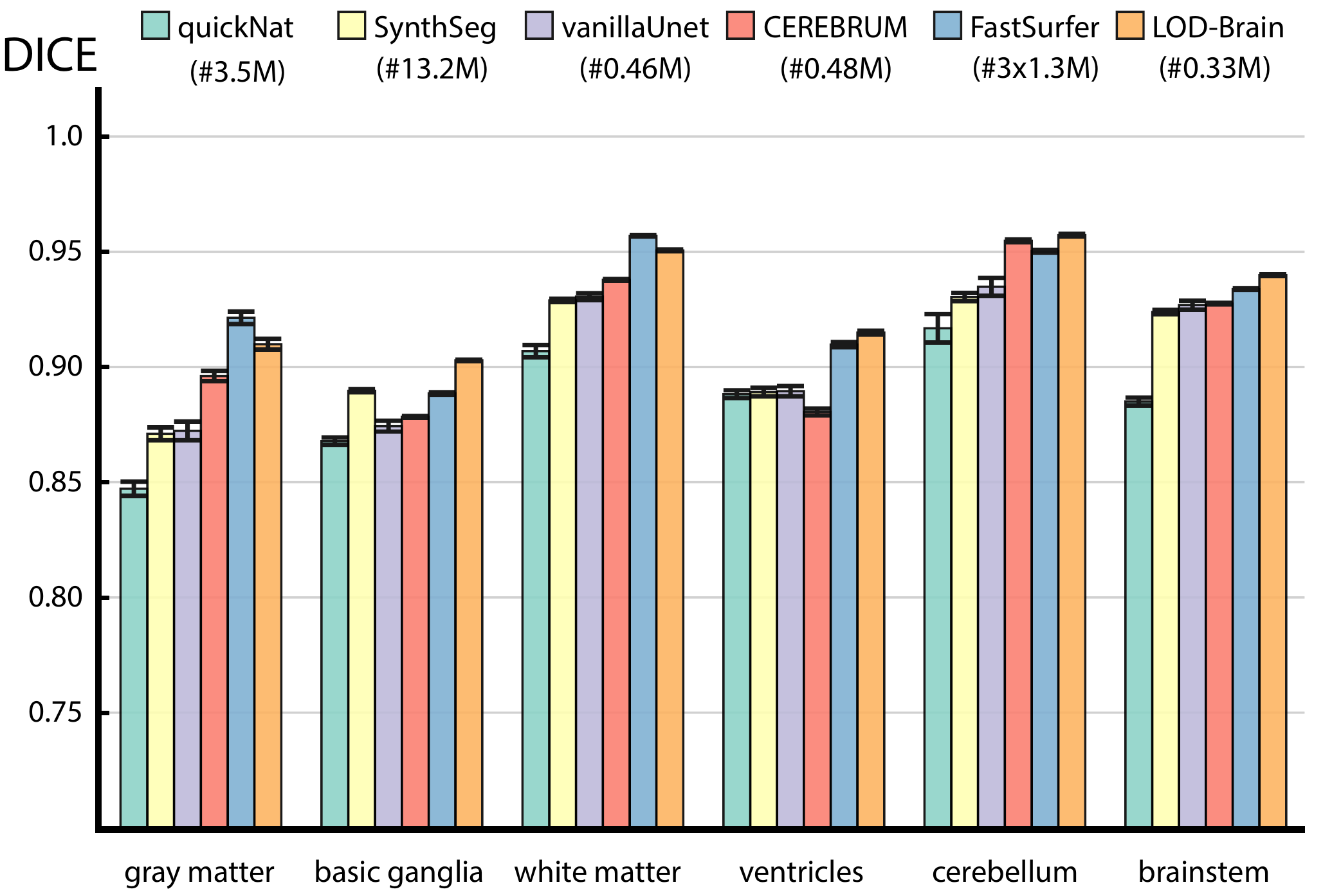}
         \caption{Results grouped by brain structure}
         \label{fig:model_comparisons_all}
     \end{subfigure}
     \begin{subfigure}[b]{\columnwidth}
         \centering
         \includegraphics[width=\textwidth]{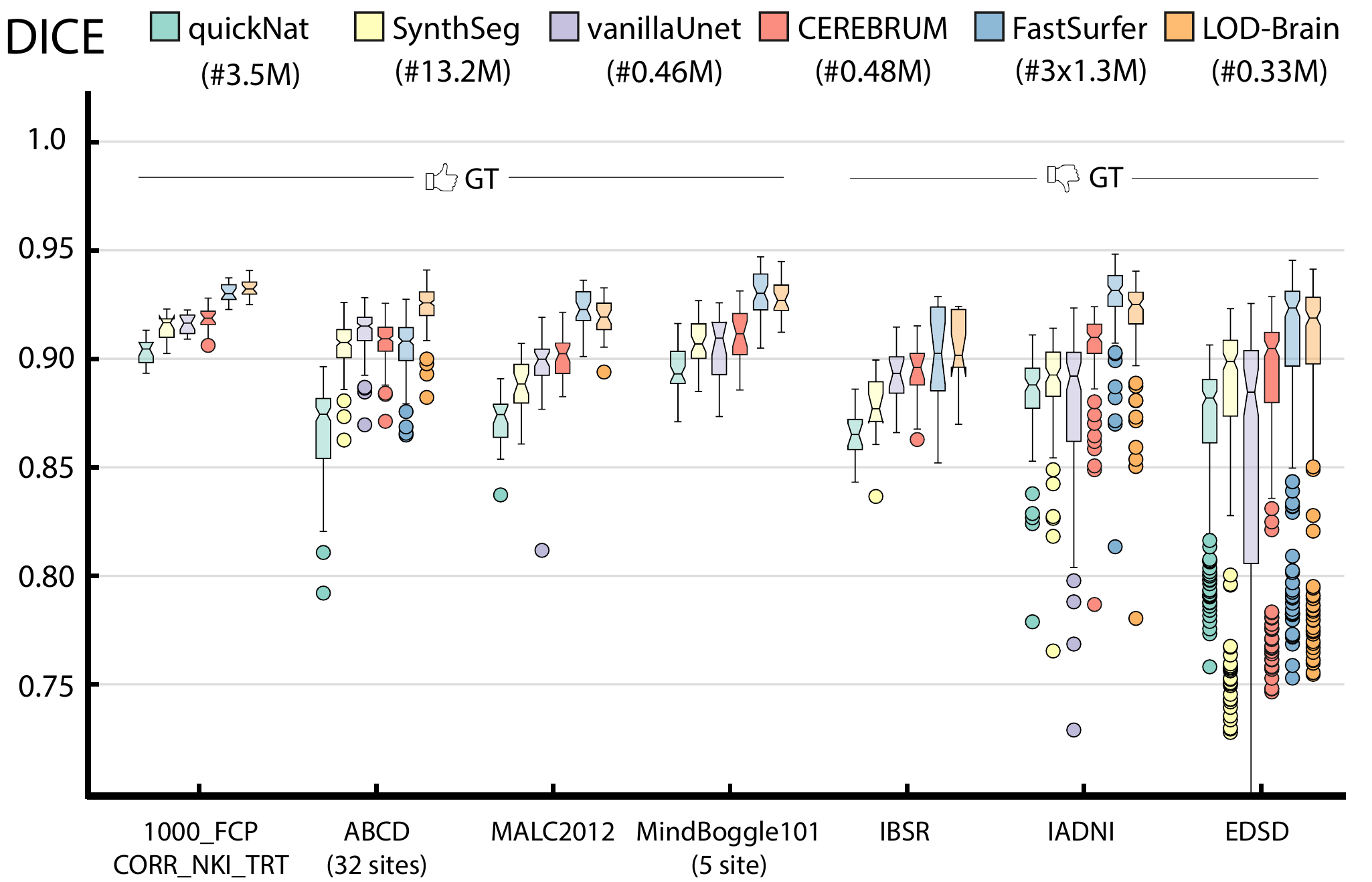}
         \caption{External datasets only}
         \label{fig:model_comparisons_EXT}
     \end{subfigure}
         \caption{
    Performance comparison: QuickNat \cite{royQuickNATFullyConvolutional2019}, SynthSeg \cite{billot2021synthseg}, 3D-UNet \cite{cicek3DUNetLearning2016b}, CEREBRUM \cite{bontempi2020cerebrum}, FastSurferCNN \cite{FastSurfer-HCE20}, and our method.
    (a) Results are computed on the test set of $5,956$ volumes, using FreeSurfer as GT reference and grouped for brain structure.
    (b) Results on external sites only, divided in acceptable vs. low-quality ground-truth.
    Numbers of parameters for each model are reported.
    }
    \label{fig:model_comparisons}
\end{figure}

A comparative assessment of our method against state-of-the-art techniques is proposed here in terms of both brain segmentation performance and model complexity. The considered benchmark methods are: QuickNat \cite{royQuickNATFullyConvolutional2019}, SynthSeg \cite{billot2021synthseg}, 3D-UNet \cite{cicek3DUNetLearning2016b}, CEREBRUM \cite{bontempi2020cerebrum}, FastSurferCNN \cite{FastSurfer-HCE20}.
Fig.~\ref{fig:model_comparisons_all} shows the obtained results on the whole testing set grouped by segmented brain structure.
Fig.~\ref{fig:model_comparisons_EXT} focuses instead on the comparative performance of different methods on external datasets only.
Obtained results highlight \texttt{LOD-Brain} as one of the most competing methods on all brain labels, as it yields the best scores in almost all target structures and on the majority of external datasets with acceptable GT.
The number of parameters for each model is also reported, highlighting \texttt{LOD-Brain} ($337,719$ parameters only) as the best overall model in terms of performance-to-complexity ratio.
It is relevant to note the high performance achieved on \href{http://abcdstudy.org/}{ABCD}, despite it includes volumes from 32 diverse scanners, previously skull-stripped and aligned to MNI152 reference space (a common situation in the field).

%

\subsection{Qualitative comparison}
\label{ssec:qualitative}
\begin{figure*}[]
    \centering
    \includegraphics[width=2.\columnwidth]{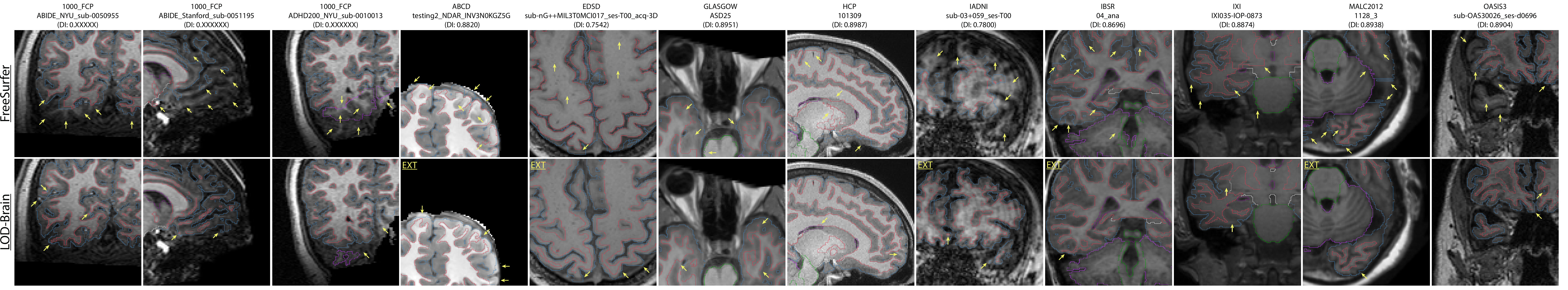}
    \caption{We select 12 volumes with the worst numerical result (max one for dataset), and display FreeSurfer (first row) and \texttt{LOD-Brain} (second row) segmentation masks overlayed to the correspondent T1w image.
    Despite low numerical performance, the segmentation masks returned by \texttt{LOD-Brain} are more accurate than Freesurfer (yellow arrows indicate errors), which is indicative of low-quality ground-truth. Full 3D volumes are displayed on the \href{https://rocknroll87q.github.io/LOD-Brain/results}{project website}.}
    \label{fig:error_analysis}
\end{figure*}

Last, in Fig.~~\ref{fig:error_analysis} we show a qualitative comparison performed on the 12 worst numerical results obtained with \texttt{LOD-Brain} (one for dataset - blue dots in Fig.~\ref{fig:complete_testing_by_site}).
We display FreeSurfer segmentation masks in the first row, and \texttt{LOD-Brain} in the second, with segmentation masks overlayed to the correspondent T1w.
Despite numerical results, which use FreeSurfer's masks as reference, the segmentation boundaries returned by \texttt{LOD-Brain} show less errors and are much smoother than those produced by FreeSurfer.

%

\section{Conclusion}
\label{sec:conclusion}

We here introduce \texttt{LOD-Brain}, a progressive level-of-detail network for training a robust brain MRI segmentation model. 
At lower levels, the network learns a strong brain prior useful to spatially identify 3D brain structures; concurrently, at higher levels, it handles site-specific and anatomical peculiarities.
Results are remarkable in terms of consistency across scanners and sites, and robust to very challenging anatomical variations. 
The proposed architecture, alongside the richness of the training dataset, leads to an automatic, fast, reliable, and off-the-shelf tool for brain MRI segmentation.
Code, model, and demo are available at the \href{https://rocknroll87q.github.io/LOD-Brain/}{project website}.

\section*{Acknowledgements}

This was supported from the EU Horizon 2020 Framework Programme for Research and Innovation under the Specific Grant Agreement No. 945539 (Human Brain Project SGA3).


 We recognise the priceless contribution made by several openly available MRI datasets: OpenNeuro (\url{https://openneuro.org/}), ABCD (\url{https://abcdstudy.org/}), Open Science Framework (\url{https://osf.io/}), the Human Connectome Project (\url{http://www.humanconnectomeproject.org/}), the NIMH Data Archive (\url{https://nda.nih.gov/}), the Open Access Series of Imaging Studies (OASIS) (\url{https://www.oasis-brains.org/}), Mindboggle101 (\url{https://mindboggle.info/data.html}), the evaluation framework for MR Brain Image Segmentation (MRBrainS) (\url{https://mrbrains18.isi.uu.nl/}), the the Amsterdam Open MRI Collection (AOMIC) (\url{https://nilab-uva.github.io/AOMIC.github.io/}), the Internet Brain Segmentation Repository (IBSR) (\url{https://www.nitrc.org/projects/ibsr}), and the great contribution provided by the International Neuroimaging Datasharing Initiative (INDI) (\url{https://fcon_1000.projects.nitrc.org/}), with many datasets, including but not limited to, the Nathan Kline Institute-Rockland Sample (NKI-RS) (\url{http://fcon_1000.projects.nitrc.org/indi/enhanced/}), the Information eXtraction from Images project (IXI) (\url{https://brain-development.org/ixi-dataset/}), the Autism Brain Imaging Data Exchange (ABIDE) (\url{http://fcon_1000.projects.nitrc.org/indi/abide/}), and the Attention Deficit Hyperactivity Disorder (ADHD) (\url{https://fcon_1000.projects.nitrc.org/indi/adhd200/}).



\bibliographystyle{apalike} 			
\bibliography{bibliography.bib}

\end{document}